\documentclass[%
 aip,
 amsmath,amssymb,
 reprint,%
]{revtex4-1}
\usepackage{xcolor}
\usepackage{graphicx}
\usepackage{dcolumn}
\usepackage{bm}
\usepackage{chemfig}
\usepackage[utf8]{inputenc}
\usepackage[T1]{fontenc}
\usepackage{mathptmx}
\usepackage{etoolbox}
\usepackage{float}  
\usepackage{physics}
\makeatletter
\def\@email#1#2{%
 \endgroup
 \patchcmd{\titleblock@produce}
  {\frontmatter@RRAPformat}
  {\frontmatter@RRAPformat{\produce@RRAP{*#1\href{mailto:#2}{#2}}}\frontmatter@RRAPformat}
  {}{}
}%
\makeatother
\begin{document}

\preprint{AIP/123-QED}

\title[Surface Excess Energy Governs the Non-Monotonic Behavior of Active Diffusivity with Activity]{
Surface Excess Energy Governs the Non-Monotonic Behavior of Active Diffusivity with Activity}
\author{A. Arango-Restrepo}
\author{J.M. Rubi}%
 \email{aarangor@unal.edu.co}
\affiliation{ 
Condensed Matter Department, University of Barcelona, 08028 Barcelona, Spain.
}%

 \homepage{https://sites.google.com/view/webjmiguelrubi/home?authuser=1}

\date{\today}

\begin{abstract}
{{Self-propulsion of particles is typically explained by phoretic mechanisms driven by externally imposed chemical, electric, or thermal gradients. In contrast, chemical reactions can enhance particle diffusion even in the absence of such external gradients. We refer to this increase as active diffusivity, often attributed to self-diffusiophoresis or self-electrophoresis, although these mechanisms alone do not fully account for experimental observations. Here, we investigate active diffusivity in catalytic Janus particles immersed in reactive media without imposed gradients. We show that interfacial reactions generate excess surface energy and sustained interfacial stresses that supplement thermal energy, enabling diffusion beyond the classical thermal limit. We consistently quantify this contribution using both dissipative and non-dissipative approaches, assuming that the aqueous bath remains near equilibrium. Our framework reproduces experimentally observed trends in diffusivity versus activity, including the non-monotonic behaviors reported in some systems, and agrees with data for nanometric Janus particles catalyzing charged substrates as well as vesicles with membrane-embedded enzymes driven by ATP hydrolysis. These results demonstrate that chemical reactions can induce and sustain surface-tension gradients and surface excess energy, providing design principles for tuning mobility in synthetic active matter.}}

\end{abstract}

\maketitle

\section{Introduction}
{Particle self-propulsion is most commonly explained by phoretic mechanisms driven by external forces or imposed gradients\cite{Anderson1989}. Stochastic thermodynamics offers a theoretical framework for understanding the dynamics of hot Brownian swimmers \cite{Cichos2010HotBP,Cichos2016HotBP}. Non-equilibrium thermodynamics combined with hydrodynamics shows that thermophoretic mobilities depend sensitively on boundary conditions,\cite{Frenkel2018_Meso,Frenkel2018unified} consistent with Faxén-type relations connecting surface-tension gradients to hydrodynamic forces\cite{Bafaluy1995Faxen}. Related studies emphasize how temperature gradients couple thermophoretic forces and flow fields\cite{Ripoll2013thermophoretically,Stark2019determining}. These works primarily address propulsion under external gradients rather than self-generated gradients in chemically reactive systems. Models of chemically induced phoresis have been proposed,\cite{Golestanian2005,Kapral2007} but they have largely focused on directed motion and phoretic velocities, leaving active diffusivity less explored. A recent study showed that applying a constant external force to randomly oriented active particles enhances diffusivity, producing an essentially constant scaling with activity \cite{Ryabov2022Enhanced}. Since this behavior is not universal, our aim is to overcome this limitations.}
\\
The enhanced mobility of chemically active nano and microparticles has been widely attributed to self-phoretic mechanisms, particularly self-diffusiophoresis, where solute concentration gradients, generated by surface chemical reactions, induce slip flows along the particle interface, resulting in propulsion \cite{Gaspar2014enzymes, popescu2016self}. This framework has explained a variety of experimental observations, especially in asymmetric systems such as catalytic Janus particles \cite{Yang2016, Kapral2014}. Nonetheless, self-diffusiophoresis alone does not fully capture the breadth of experimentally observed diffusivity behaviors, especially regarding their nonlinear dependence on reaction rates \cite{Zhou2018, Ji2019}. Self-electrophoresis has been introduced as a complementary mechanism \cite{Kapral2013, Kuron2018}, yet a unified understanding remains lacking.
\\
A thermodynamically grounded yet often underappreciated mechanism is self-thermophoresis, arising from temperature gradients produced by exothermic surface reactions \cite{Qin2017_self-thermo, Kapral2013}. While extensively studied in externally forced systems \cite{Jiang2010, Carles2020}, its contribution as a self-generated driving force in active particles is less explored. Moreover, existing models tend to treat diffusiophoresis, electrophoresis, and thermophoresis in isolation, overlooking the possible couplings among them and their collective impact on particle dynamics.
\\
At a fundamental level, self-phoretic motion arises from thermodynamic gradients, such as concentration, temperature, or electric potential of the solute, generated by chemical reactions on the surface of the particle \cite{Popescu2022, Kapral2013}. These gradients induce slip flows that drive motion but also modify the surface excess energy, defined as the energy stored at the interface due to local deviations from ideality. Importantly, the formation and maintenance of such gradients lead to irreversible processes that generate entropy at the interface, sustaining a non-equilibrium steady state \cite{Arango-Restrepo2024}. This entropy production reflects ongoing energy dissipation, which plays a direct role in enhancing particle mobility.
\\
This perspective shifts the focus from purely mechanical descriptions of propulsion to a thermodynamic interpretation, where surface tension gradients, energy fluxes, and entropy generation collectively determine the mobility of active particles. Crucially, these interfacial thermodynamic variables are accessible through measurable quantities and can be used to develop predictive, quantitative models of active diffusivity. As we show in this work, incorporating both non-dissipative and dissipative contributions to the surface excess energy provides a comprehensive understanding of how interfacial processes govern enhanced particle transport.
\\
{In this work, we develop a thermodynamic framework to explain the enhanced diffusivity of Janus catalytic particles, observed in experiments, through explicit analysis of the nonequilibrium processes taking place in the particle-fluid interface. Our goal is to unify the main self-phoretic mechanisms (diffusiophoresis, electrophoresis, and thermophoresis) through their connection to excess surface energy. We calculate the excess surface energy generated by catalytic reactions using two complementary approaches: a non-dissipative one, which captures how reaction-induced gradients modify the interfacial energy, and a dissipative one, which considers entropy generation and energy dissipation. Both perspectives reveal how catalytic activity and thermal energy enhance diffusion beyond the classical thermal limit when considering that the vast aqueous media is at thermal equilibrium. From the resulting expression for diffusivity, we recover the experimentally observed non-monotonic dependence of diffusivity on reaction rate. To validate the framework, we analyze two experimentally studied systems: nanometric Janus particles using charged salts as substrates \cite{Qin2017_self-thermo}, and phospholipid vesicles with integrated enzymes that hydrolyze ATP \cite{Ghosh2019}. Our findings highlight surface tension gradients and interfacial entropy production, through excess surface energy, as central regulators of active diffusivity and provide a solid physical basis for the design of synthetic active matter.}
\\
The paper is organized as follows. Section II presents the system and describes the theoretical framework of active diffusivity, along with the estimation of excess surface energy from dissipative and non-dissipative perspectives. Section III presents the model based on conservation equations and on the calculation of entropy production. In Section IV, we derive analytical expressions for surface energy excess using both approaches. Section V contains our results, including a comparison with experimental data, and highlights the nonlinear and nonmonotonic dependence of active diffusivity on activity (reaction rate). Finally, Section VI summarizes our main conclusions.

\section{Theory}

\subsection{System}

\begin{figure}[h!]\centering
\centering
    \includegraphics[width=7.25cm]{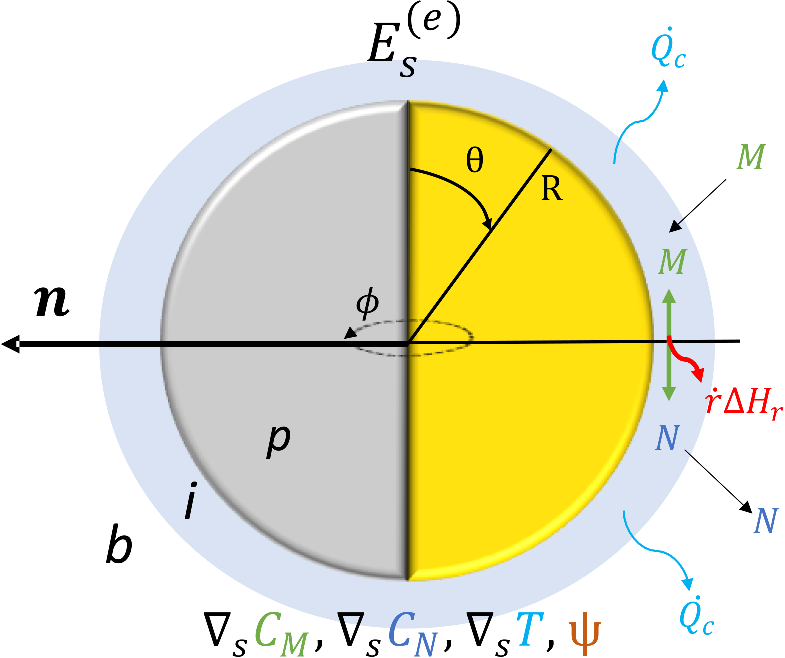}
\caption{Illustration of a catalytic Janus particle undergoing a first-order reaction at its interface, where substrate $M$ is converted into product $N$ at rate $\dot{r}$ producing heat $\dot{r}\Delta H_r$, and thereby inducing an excess surface energy $E_{s}^{(e)}$. Here, the particle is depicted with an orientation $\mathbf{n}$, extending from the catalytic (golden color) to the non-catalytic side (grey color). The interface region $i$ is located between the inner section of the particle $p$ and the surrounding fluid $b$. From the interface, the heat flux $\dot{Q}_c$ is transferred to the fluid, while $M$ is being absorbed from the fluid. The surface of the particle is parametrized by the polar $\theta$ and azimuthal $\phi$ angles. }
\label{Fig_Janus}
\end{figure}

We consider a catalytic Janus particle where a chemical reaction on one side of the surface generates asymmetric concentration and temperature fields. On the surface of the particle, an irreversible chemical reaction takes place, which converts a substrate $M$ into products $N$. This process occurs in a medium without external flow and interparticle interactions. At the interface ($i$) between the particle ($p$) and the surrounding fluid ($b$), all chemical species are present. Figure \ref{Fig_Janus} depicts a spherical catalytic Janus particle and its surrounding environment. The active diffusion is primarily governed by the surface excess energy, \( E_{s}^{(e)} \), which results from surface concentration and temperature gradients (\(\nabla_S C_M\), \(\nabla_S C_N\), \(\nabla_S T\)), the electrostatic potential \(\psi\), and surface entropy \( S \). These variations originate from the reaction rate \( \dot{r} \) and the heat generated by the reaction, \( \dot{r} \Delta H_r \), within the catalytic region of the particle.

\subsection{Active diffusivity}

We consider a spherical particle of mass \( m \) and moment of inertia \( I \), whose orientation is described by a unit vector \( \mathbf{n}\). The underdamped Langevin equations governing the translational velocity \( \mathbf{v}(t) \) and the angular velocity \( \boldsymbol{\omega}(t) \) are:
\begin{equation}
\label{lang_t}
    m \dot{\mathbf{v}} = -\xi_t \mathbf{v} + \mathbf{F}_{\text{ph}} +\mathbf{F}_t(t)
\end{equation}
\begin{equation}
    I \dot{\boldsymbol{\omega}} = -\xi_r \boldsymbol{\omega} + \mathbf{T}_{\text{ph}} + \mathbf{T}_r(t)
\end{equation}
where \( \xi_t \) and \( \xi_r \) are the translational and rotational friction coefficients, respectively, \( \mathbf{F}_{\text{ph}}=\xi_t b \mathbf{n} \) is the phoretic (active) force with $\mathbf{n}$ the orientation vector and $b$ to be determined, whereas \( \mathbf{T}_{\text{ph}} \) is the phoretic (active) torque {fulfilling \( \mathbf{T}_{\mathbf{ph}}\otimes\mathbf{n} = (\mathbf{I}-\mathbf{n}\mathbf{n})\cdot\nabla_{\mathbf{n}}\psi \), with $\psi = E_{s}^{(e)}(1-\mathbf{n}\cdot\mathbf{m})^2$ the orientation potential and $\mathbf{m}$ the external field unitary director vector\cite{Arango-Restrepo2025}}. The random force for translational velocities \( \mathbf{F}_t \) and the random torque for angular velocity \( \mathbf{T}_r \), are Gaussian white noises that fulfill the fluctuation-dissipation theorem: $\langle \mathbf{F}_t(t) \mathbf{F}_t(t') \rangle = 6\xi_tk_BT^{(b)}\mathbf{I}\delta(t - t')$ and $\langle \mathbf{T}_r(t) \mathbf{T}_r(t') \rangle = 6\xi_rk_BT^{(b)}\mathbf{I}\delta(t - t')$ {with $T^{(b)}$ the bulk/bath temperature}.
\\
The orientation vector \( \mathbf{n} \) evolves according to:
\begin{equation}
\label{omega}
    \dot{\mathbf{n}} = \boldsymbol{\omega} \times \mathbf{n}
\end{equation}
in which \( \norm{\mathbf{n}} = 1 \) for all times. In the absence of external fields or gradients that could align the particle\cite{Arango-Restrepo2024}, the phoretic torque is negligible compared to the rotational noise, thus $\mathbf{n}$ is random, so its correlation for long times is $\langle \mathbf{n}(t) \mathbf{n}(t') \rangle \approx \mathbf{I}\delta(t - t')$ (see Appendix A). 
\\
We can then see that the random phoretic force fulfills the fluctuation-dissipation theorem $\langle \mathbf{F}_{ph}(t) \mathbf{F}_{ph}(t')\rangle = 6\xi_tB\mathbf{I}\delta(t-t') $, with $B$ the phoretic energy of the particle-fluid interphase. Solving Eq.(\ref{lang_t}) (see Appendix B) for long times, we obtain the mean kinetic energy of the particle as a function of the thermal energy of the solvent and the resulting magnitude of the excess surface energy $|E_{s}^{(e)}|$ on the particle due to the reaction
 \begin{equation}
  \frac{m}{2}\langle{v}(t)^2 \rangle_{t\to\infty}  =  \left|E_{s}^{(e)}\right| + k_BT^{(b)} 
\end{equation}
Notice that in equilibrium, the fluctuation-dissipation theorem (FDT) ensures that fluctuations and dissipation are balanced, with diffusivity determined solely by the thermal energy of the solvent via the equipartition theorem. However, for active particles, additional energy—originating from chemical reactions or surface activity—injects fluctuations beyond those allowed by thermal equilibrium. This leads to a violation of the FDT: the particle experiences enhanced movement and mean-squared velocity without a corresponding change in friction. Mechanically, this means the diffusivity reflects not just thermal fluctuations but also excess surface energy, effectively raising the energy available per degree of freedom and resulting in faster motion even in the absence of external forces. 
\\
Considering that the mean displacement of the particle is given by $\Delta x(t) = \int_0^t v(t)dt$ with $v(t)$ given in Eq.(\ref{v_sln}) \cite{zwanzig2001nonequilibrium}, we can obtain the mean square displacement $\langle \Delta x^2\rangle(t)$ for long times from Eq.(\ref{lang_t})\cite{Arango-Restrepo2024}
\begin{equation}
    \langle \Delta x(t)^2\rangle_{t\gg t_0} = 2\left( \frac{k_{B}T^{(b)} }{\xi_{t}}+  \frac{\left|E^{(e)}_s\right| }{\xi_{t}}  \right)t
\end{equation}
Therefore, the active diffusivity is given by
\begin{equation}
\label{Diff}
  D =D_{0}\left(1 +  \frac{\left|E^{(e)}_s\right| }{k_{B}T^{(b)}}\right) 
\end{equation}
where $D_0$ is the diffusivity of the particle in the absence of reaction.

\subsection{Excess surface energy}

The surface excess energy of an active particle arises from local mass and heat fluxes from a surface reaction that also generate entropy at the interface. The chemical reaction then drives energy input at the surface, increasing fluctuations and enhancing the particle's mobility. As a result, the surface becomes a source of nonequilibrium energy, linking fluxes (or gradients) and entropy production directly to the particle’s effective diffusivity. Therefore, in this subsection we will define the surface excess energy from a non-dissipative (surface tension gradient) and dissipative (entropy production) perspectives.

\subsubsection{Non-dissipative approach}

To determine the excess surface energy, we analyze an infinitesimal local change in the interfacial energy, $U = Au$, where $A$ denotes the surface area of the particle, 
\begin{equation}
    dU = \delta Q + \sum_i \mu_idN_i + \gamma dA
\end{equation}
Here $\delta Q$ represents the reversible heat exchanged between the interface and the fluid, including the heat released by a chemical reaction taking place at $dA$, $\mu_i$ denotes the chemical potential of component $i$, and $N_i$ is the number of moles of component $i$ at $dA$. 
\\
Defining the specific surface excess energy $e_{s}^{(e)}$ as the difference between the internal energy per area $u$ and the surface tension $\gamma$: $e_{s}^{(e)} = u-\gamma$, and re-writing the energy balance, we obtain the differential variation of the specific surface energy excess
\begin{equation}
\label{e_e}
    de_{s}^{(e)} = \delta q + \sum_i \mu_idC_i - d\gamma
\end{equation}
with $q = Q/A$ and $C_i = N_i/A$. For a non-reactive system composed of a pure fluid, we recover the well-known result $e_{s}^{(e)} = -T\,\partial\gamma/\partial T$, as previously reported in Ref.~\cite{Hirschfelder1954}. 
\\
Focusing on the heat and work required to increase the excess surface energy in a reactive system, we consider the following:
\\
1) The surface heat variation is determined by the reversible heat locally produced (or consumed) by an exothermic (or endothermic) reaction, given approximately by $\delta q \approx -\Delta H^{0}\,d\xi$, with $\xi$ the extent of reaction.
\\
2) The chemical potential is expressed as $\mu_i = \mu_i^{0} + R_g T \ln(f_i x_i) + z_i F \psi$, where $R_g$ is the gas constant, $\mu_i^{0}$ is the standard chemical potential, $f_i$ the activity coefficient (accounting for non-ideality), $z_i$ the charge of component $i$, $x_i$ the molar fraction, $F$ the faraday constant and $\psi$ the electric potential.
\\
3) The variation of surface tension depends on local changes in temperature and concentration: $d\gamma = \gamma_T\,dT + \sum_i \gamma_{C_i}\,dC_i$.
\\
Taking the surface differential, we obtain the local surface gradient of the specific surface energy excess:
\begin{equation}
    \begin{split}
        \nabla_s e_{s}^{(e)} = & -\Delta H^{0} \nabla_s\xi + \sum_i \left( \mu_i^{0} + R_g T \ln(f_i x_i) \right)\nabla_s C_i \\ & + F\psi\sum_i z_i \nabla_s C_i - \gamma_T \nabla_sT - \sum_i\gamma_{C_i}\nabla_s C_i
    \end{split}
\end{equation}
Defining the surface energy excess $E_s^{(e)} = Ae_{s}^{(e)}$ and integrating along the surface, we obtain
\begin{equation}
\label{Excess}
    \begin{split}
        E_{s}^{(e)} = & -A\Delta H^{0} \left\langle \nabla_s\xi\right\rangle + A\sum_i \left( \mu_i^{0} + R_g   \left\langle T\ln(f_i x_i) \right)\nabla_s C_i\right\rangle \\ & + AF\sum_i z_i \left\langle\psi\nabla_s C_i\right\rangle - A\gamma_T \left\langle\nabla_sT\right\rangle - A\sum_i\gamma_{C_i}\left\langle\nabla_s C_i\right\rangle
    \end{split}
\end{equation}
in which we have defined $\int_{S'} ... dS' = \langle...\rangle$. As a consequence, only active particles that create asymmetric distributions of temperature or concentrations, due to an asymmetric reaction rate, develop excess energy, leading to active diffusion. 
\\
Furthermore, the dependence on surface gradients implies a coupling with reaction-induced surface mass and heat fluxes. This leads to the interpretation of surface energy excess, and consequently, active particle diffusivity, as being driven by the non-dissipative currents within the overall dissipative process \cite{maes2017non,MAES20201}.

\subsubsection{Dissipative approach}

From the previous observation, we wonder whether the total entropy change might also encode information about a dissipative component of the surface energy. In this subsection, we explore this possibility by examining how entropy production at the interfaces contributes to energy conversion and transport, aiming to characterize a thermodynamically consistent framework for surface-driven activity.
\\
To do this, we write the differential change of the excess entropy at the surface
\begin{equation}
    \delta S = -R\sum_i \ln{x_i}dN_i + \delta\Sigma
\end{equation}
in which $\delta\Sigma$ is the irreversible change of the entropy at the surface. Taking the time derivative and defining the entropic change of the surface times the temperature as the dissipative surface energy $E_{s}^{(d)}$, we have
\begin{equation}
\label{Diss_Excess}
    \frac{E_{s}^{(d)}}{\tau} =  -R_g\sum_i \left\langle T\dot{R}_i\ln{x_i}\right\rangle + \left\langle T\sigma\right\rangle
\end{equation}
with $\dot{R}_i$ the local reaction rate, $\sigma$ the local entropy production rate and $\tau$ the characteristic time of the process. 
\\
Surface dissipative energy links the entropic cost associated with maintaining concentration gradients that emerge due to a chemical reaction with the intrinsic production of entropy at the interface. It is noteworthy that the activity driven by the surface reaction is thermodynamically sustained not only by the contribution of chemical energy, but also by the excess entropy derived from mixing and energy dissipation, which continuously reorganise matter and energy at the interface.
\\
Finally, given the dissipative surface excess energy, the active diffusivity based on the dissipative approach is
\begin{equation}
 \label{D_d}
     D^{(d)} = D_0\left( 1 + \frac{\left| E_{s}^{(d)} \right|}{k_BT}\right)
\end{equation}
This perspective offers an alternative approach to exploring dissipation's role in transport properties' non-trivial behavior. This leads to an interpretation of excess surface dissipative energy—and thus the active diffusivity—as arising solely from the irreversible components of the dissipative process cite\cite{Arango-Restrepo2021}.

\section{Model}

We consider {as our studied system} the interface between a catalytic Janus particle and the surrounding fluid (Fig.\ref{Fig_Janus}). The system operates in a non-equilibrium steady state, under the assumption that local thermodynamic equilibrium (LTE) holds at the interface. Our model involves a first-order surface reaction characterized by a kinetic constant that is independent of temperature. This approximation is valid when the interfacial thermal conductivity is high, the reaction enthalpy is low, or the temperature variations along the interface are minimal, though not entirely absent. A similar assumption is made for all parameters, which are considered constant due to the small variations in temperature and concentration, and the typically dilute nature of the system.
\\
{This assumption is justified because the suspension is dilute and immersed in a large thermal bath, so the bath temperature remains effectively constant and the surrounding fluid acts as an equilibrium reservoir. Under these conditions, the particle–fluid interface remains in local equilibrium throughout the transient relaxation process, which ultimately converges to equilibrium once the chemical fuel is consumed. These conditions have been extensively validated through nonequilibrium molecular simulations \cite{Bjorn1995criteria,Bjorn2023_LET,Bjorn2025_LET_Diff}, which demonstrate that LTE holds reliably under such circumstances.}

\subsection{Balance equations at the interface}

The key surface quantities can be obtained by solving the conservation laws. The substrate concentration at the interphase between an AP and the bulk fulfills the substrate mass conservation equation
\begin{equation}
\label{EQ_Cons}
    \mathbf{v}_{S}\cdot \nabla_s C_M= -\nabla_s\cdot\mathbf{J}_M - k_{r}C_{M} \Theta(\Omega_0-\Omega) - U(C_{M}-C_{M}^{(b)})
\end{equation}
On the left-hand side, the term represents the advective contribution to substrate transport along the interface. On the right-hand side, the first term captures the diffusive contribution driven by concentration, temperature, and electric field gradients; the second term accounts for the chemical reaction contribution; and the final term describes mass transfer in the radial direction. Here $\mathbf{v}_{S}$ is the surface velocity of the interface, $k_{r}$ is the reaction constant, $U$ stands for the mass transfer coefficient, $C_{M}^{(b)}$ is the bulk substrate concentration and $\Omega_0$ the borderline between both sides of the particle. The diffusive current is given by \cite{de2013non}
\begin{equation}
\label{J_d}
    \mathbf{J}_M = -D_M\nabla_sC_M - D_M\mathcal{S}_sC_M\nabla_MT - D_M\frac{q_0z_MC_M}{k_BT}\nabla_s\psi 
\end{equation}
in which $D_M$ denotes the diffusivity of the substrate at the interface, $\mathcal{S}_M$ the substrate Soret coefficient and $q_0$ the electron charge per mole. The product concentration balance equation is
\begin{equation}
\label{EQ_ConsN}
    \mathbf{v}_{S}\cdot \nabla_s C_N= -\nabla_s\cdot\mathbf{J}_N + k_{r}C_{M} \Theta(\Omega_0-\Omega) - U(C_{N}-C_{N}^{(b)})
\end{equation}
with $\mathbf{J}_N$ similar to $\mathbf{J}_M$ considering the transport properties of the product instead of the substrate.
 \\
Since the reaction kinetics depends only on the substrate concentration, the energy balance at the interface is:
\begin{equation}
\label{EQ_ConsT}
\mathbf{v}_{S}\cdot \nabla_s T= \kappa\nabla_s^2T - k_{r}C_{M}\Delta H_{r} \Theta(\Omega_0-\Omega) - U_q(T-T^{(b)})
\end{equation}
The left-hand side term corresponds to the advective contribution to the heat transport along the interface. On the right-hand side, the first term accounts for the diffusive contribution, the second represents the heat produced (or consumed) due to the exothermic (endothermic) chemical reaction, and the last term corresponds to the heat transfer in the radial direction. $\kappa$ denotes the heat diffusivity constant at the interface, $\Delta H_{r}$ is the reaction enthalpy (negative for exothermic, and positive for endothermic reactions), $U_q$ stands for the heat transfer coefficient, and $T^{(b)}$ is the bulk temperature. {Note that we explicitly include tangential and normal transport contributions, with mass adsorption/desorption (last terms in Eqs.(\ref{EQ_Cons}) and (\ref{EQ_ConsN})) and heat flow (Eq.(\ref{EQ_ConsT})) occurring in the direction normal to the interface.}
\\
{The electrostatic potential $\psi$ at a particle–fluid interface can be obtained from the Poisson equation
\begin{equation}
\nabla_s^2\psi = -\frac{q_0}{\varepsilon}\sum_i z_i C_i
\end{equation}
with $\varepsilon$ the permittivity of the medium. We have assumed a homogeneous charge distribution in the surrounding fluid, consistent with the dilute limit for Janus particles, where interparticle interactions are negligible. Within this approximation, the ions in solution contribute only a constant offset to the potential, without significantly modifying its spatial structure. At low ion concentrations, ionic interactions are weak, and the solvent can be treated as a continuous medium with nearly uniform ion distribution, as described by Debye–Hückel theory.}

\subsection{Entropy production at the interface}

The entropy production rate at the particle surface is given by \cite{BEDEAUX1976, Gaspard_2019, Gaspard2020am, ARANGORESTREPO2018}
 \begin{equation}
 \label{simga_O}
 \begin{split}
          \sigma = & -\frac{1}{T}\boldsymbol{\Pi}_s:\nabla_s\mathbf{v}_{s} - \frac{1}{T}\mathbf{J}_M\cdot \nabla_{s}\mu_M  -\frac{1}{T}\mathbf{J}_N\cdot\nabla_{s} \mu_N  \\&-\frac{1}{T^2}\mathbf{J}_q\cdot \nabla_{s}T - \frac{1}{T}\int_\nu J_r \frac{\partial \mu}{\partial \nu}d\nu
 \end{split}
 \end{equation}
in which the dissipative forces are the surface gradient of the slip velocity, $\nabla_{s}\mathbf{v}_{s}$\cite{BEDEAUX1976}, chemical potential surface gradients of substrate, $\nabla_{s} \mu_M$, and product, $\nabla_{s} \mu_N$, temperature surface gradient, $\nabla_{s} T$\cite{Gaspard_2019,Gaspard2020am} and $\frac{\partial \mu}{\partial \nu}$ the derivative of the chemical potential $\mu$ of the substrate along the reaction coordinate $\nu$\cite{ARANGORESTREPO2018}. Furthermore, entropy production depends on the diffusive fluxes of the substrate ($\mathbf{J}_{M}$) and product ($\mathbf{J}_{N}$) currents, the heat flux ($\mathbf{J}_{q}$) and reactive flux in $\nu$-space ($J_r$).
\\
\begin{table*}
\caption{\label{tab:table1}Analytic solutions of the balance equations}
\begin{ruledtabular}
\begin{tabular}{ccc}
 Variable & Catalytic side & Non-catalytic side \\ \hline
 $C_M/C_0$&$\frac{\beta^2}{\alpha^2 + \beta^2} + 2k^{+}\cosh(\sqrt{\alpha^2 + \beta^2}\phi)$& $2k^{-}\exp(-\beta\pi)\cosh(\beta(\pi-\phi)) $\\
 $C_N/C_0$&$ 4\frac{\alpha^2 k^{+}}{\alpha^2 + \beta^2}\cosh(\sqrt{\alpha^2+\beta^2}\phi) + m^{+}\cosh(\beta\phi) $&$2m^{-}\exp{(-\beta\pi)}\cosh(\beta(\pi-\phi))$\\
 $T/T_0$& $\!\begin{aligned}[t]
    &\frac{k^{+}\lambda^2}{\omega^2}\cosh{(\sqrt{\alpha^2+\beta^2}\phi)} + n^{+}\cosh{(\omega\phi)}  \\
    & + \frac{\lambda^2\beta^2}{\omega^2(\alpha^2+2\beta^2)}(1+2\alpha^2) + 1   
    \end{aligned}$ &$ 2n^{-}\exp{(-\omega\pi)}\cosh{(\omega(\pi-\phi))} + 1$\\
 $-\psi/\psi_0\xi^2$& $\frac{\beta^2}{\alpha^2+\beta^2}(\phi^2-\pi\phi-w_1) - \frac{2k^+}{\alpha^2+\beta^2}\cosh{(\sqrt{\alpha^2+\beta^2}\phi)}$\footnote{$w_1=\frac{\beta^2\left(\left(1 - \sqrt{(\alpha/\beta)^2+1}\right)^{-2}-1\right)-\alpha^2}{\beta^2(\alpha^2+\beta^2)}$} &$2\frac{k^-}{\beta^2}\exp{(-\beta\pi)}\cosh{(\beta(\pi-\phi))}$\\
\end{tabular}
\end{ruledtabular}
\end{table*}
The mass fluxes were previously defined in Eq.~(\ref{J_d}), while the heat flux at the interface is approximated by $\mathbf{J}_{q} = \kappa \nabla_s T$, where $\kappa$ is the interfacial thermal conductivity. The local contribution to the entropy production due to the chemical reaction is estimated as $\frac{R_g}{C_0}k_r C_M^2$~\cite{ARANGORESTREPO2018}. The surface velocity corresponds to the phoretic slip velocity, given by $\xi_t \mathbf{v}_s = \int_{S'} \nabla_s \gamma \, dS'$, where $\xi_t$ is a mobility coefficient and $\gamma$ the surface tension \cite{Arango-Restrepo2024}. The surface stress tensor for a Newtonian interface is expressed as \cite{SCRIVEN1960,BEDEAUX1976}:
\begin{equation}
\boldsymbol{\Pi}_s = \eta_s \left( \nabla_s \mathbf{v}_s + (\nabla_s \mathbf{v}_s)^\top \right) + (\eta_d - \eta_s)(\nabla_s \cdot \mathbf{v}_s)\, \mathbf{I}_s,
\end{equation}
where $\eta_s$ and $\eta_d$ are the surface shear and dilatational viscosities, respectively, and $\mathbf{I}_s$ is the identity tensor projected onto the surface.

\section{Analytic expressions for the surface excess energy}

Our goal is to explicitly demonstrate the dependence of active diffusivity (and surface excess energy) on the mean substrate concentration and, consequently, on the mean reaction rate. We will obtain an analytical solution to the previous equations.  We first define the catalytic zone as \( -\pi/2 \leq \phi \leq \pi/2 \). Secondly, we assume field symmetry at \( \theta = \pi/2 \), which implies that the variables primarily depend on \( \phi \). Third, we express the dimensionless balance equations as presented in Ref. \cite{Arango-Restrepo2024} and Appendix C. Dimensional analysis shows that the thermodiffusion and electrodiffusion of the substrate and product on the surface (Eq. \ref{J_d}) can be neglected when the diffusivities of both species are similar, their charges are exactly opposite, and the Soret coefficient is lower than 1. 
\\
In Table \ref{tab:table1}, we present the solutions for the balance equations, which explicitly depend on the dimensionless numbers: $\alpha^2 = k_rR^2/D_M$, $\beta^2 = UR^2/D_M$, $\lambda^2 = R^2|\Delta H_r|k_rC_0/\kappa T_0$, $\omega^2 = U_qR/\kappa$, and $\xi^2$, quantifying reaction-diffusion, adsorption-diffusion, heat production-conduction, heat release-conduction, and concentration-permittivity effects, respectively. The parameters $k^+$, $k^-$, $m^+$, $m^-$, $n^+$ and $n^-$ are determined by applying the continuity condition at the boundary at $\phi = \pi/2$. Finally, $C_0$ and $T_0$ denote the bulk concentration of the substrate and temperature, while $\psi_0 = k_B T_0/q_0$ represents the characteristic electrostatic potential.  
\\
The surface excess energy for the previous considerations is then expressed as
\begin{equation}
\label{Excess1}
\begin{split}
    E_s^{(e)} = & A\epsilon(\Delta H_r +\Delta\mu_r^0  ) \int_0^\pi \frac{\partial C_M}{\partial \phi}\sin{\phi} d\phi \\ & + A\epsilon R_g\sum_i \int_0^\pi T\ln{C_i}\frac{\partial C_i}{\partial \phi}sin\phi d\phi  \\ & + FA\epsilon\sum_i z_i \int_0^\pi \psi\frac{\partial C_i} {\partial\phi}\sin{\phi}d\phi \\ & - A\sum_i \gamma_{C_i} \int_0^\pi \frac{\partial C_i}{\partial \phi}\sin{\phi} d\phi \\ & - A\gamma_T \int_0^\pi \frac{\partial T}{\partial \phi}\sin{\phi} d\phi
\end{split}
\end{equation}
in which it was assumed a diluted substrate, and where $\epsilon$ is the interface thickness.  
\\
\begin{table}
\caption{\label{tab:table2}Surface energy contributions}
\begin{ruledtabular}
\begin{tabular}{ccc}
 Surface energy source  & Result \\ \hline
 $E_{s,r}^{(e)}$& $-A\epsilon(\Delta H_r +\Delta\mu_r^0  )\frac{1-\sqrt{(\alpha/\beta)^2+1}}{1+\sqrt{(\alpha/\beta)^2+1}}\langle C_M\rangle$ \\
 $E_{s,S}^{(e)}$&$A\epsilon R_gT_0 \ln{\left(\langle \hat{C}_M\rangle(1-w_0)\right)}(\alpha^2+\beta^2)\langle C_M\rangle$\footnote{For $w_0 = \left(1 + \sqrt{(\alpha/\beta)^2+1} \right)^{-1}$}\\
 $E_{s,\psi}^{(e)}$& $\epsilon FA\frac{k_BT}{q_0}\xi^2 \left[ \frac{\langle \hat{C}_M\rangle z_M}{w_0\beta} -\frac{\alpha^2}{\alpha^2+\beta^2}z_N  \right]\langle C_M\rangle$\\
$E_{s,\gamma_{C_i}}^{(e)}$&$A\left(\gamma_{C_M}\frac{1-\sqrt{(\alpha/\beta)^2+1}}{1+\sqrt{(\alpha/\beta)^2+1}} + \gamma_{C_N} \frac{\alpha^2}{\alpha^2+\beta^2}\right)\langle C_M\rangle$\\
$E_{s,\gamma_{T}}^{(e)}$&$A\gamma_T \frac{T_0}{C_0}  w_0 \frac{\lambda^2}{\omega^2}(\alpha^2+2\beta^2)\langle C_M\rangle$ \\
\end{tabular}
\end{ruledtabular}
\end{table}
In Table \ref{tab:table2}, we present the solution for each term of the surface excess energy for \( \alpha^2 \ll 1 \) and \( \beta^2 \ll 1 \). The first term in Eq.~(\ref{Excess1}), \( E_{s,r}^{(e)} \) (see Table~\ref{tab:table2}), represents the enthalpic and free energy effect due to chemical reaction. The second term in Eq.~(\ref{Excess1}), \( E_{s,S}^{(e)} \), corresponds to the entropic contribution arising from spatial variations in surface concentration. The third term, \( E_{s,\psi}^{(e)} \), represents the self-electrophoretic part. The fourth term, \( E_{s,\gamma_{C_i}}^{(e)} \), accounts for the self-diffusiophoretic effect, while the last term, \( E_{s,\gamma_T}^{(e)} \), describes the self-thermophoretic contribution. The magnitude of the surface excess energy is the absolute value of the sum of the components:  $\left|E_s^{(e)}\right| = \left|E_{s,r}^{(e)} + E_{s,S}^{(e)} + E_{s,\psi}^{(e)} + E_{s,\gamma_{C_i}}^{(e)} + E_{s,\gamma_{T}}^{(e)}\right|$.
\\
Our study reveals that the surface excess energy is proportional to the average surface substrate concentration, \( \langle C_M \rangle = \frac{\beta^2}{\alpha^2 + \beta^2} \), which is directly related to the average reaction rate, \( \langle J_r \rangle = k_r \langle C_M \rangle \). Furthermore, when the electrostatic contribution dominates, the excess energy scales with the square of the average substrate concentration, i.e., \( E_s^{(e)} \propto \langle C_M \rangle^2 \).
\\
It is important to emphasize that the excess energy is strongly influenced by the ratio between reaction and absorption \(\alpha/\beta\), and between heat production and cooling rate \(\lambda/\omega\). When absorption dominates, i.e., $\beta \gg \alpha$, self-thermophoresis becomes the primary source of surface excess energy. In contrast, in a reaction-dominated regime, i.e., $\beta \ll \alpha$, self-diffusiophoresis, self-electrophoresis, and enthalpic-free energy variations play the dominant role. For micrometric particles, $\alpha^2$ and $\beta^2$ approach to one, making entropic effects increasingly significant. Additionally, for highly exothermic reactions with poor heat dissipation, i.e., $\lambda^2 \sim \omega^2$, self-thermophoresis is amplified and may become the main contributor to surface excess energy.
\\
To confirm the assumption of symmetric fields around $\theta = \pi/2$, we numerically solve the surface balance equations. Additionally, we analyze how the contributions to surface excess energy depend on the mean substrate concentration. To ensure the generality of our results, we analyzed the variations in the dimensionless quantities.

\begin{figure}[h!]
    \centering
    \includegraphics[width=1.025\linewidth]{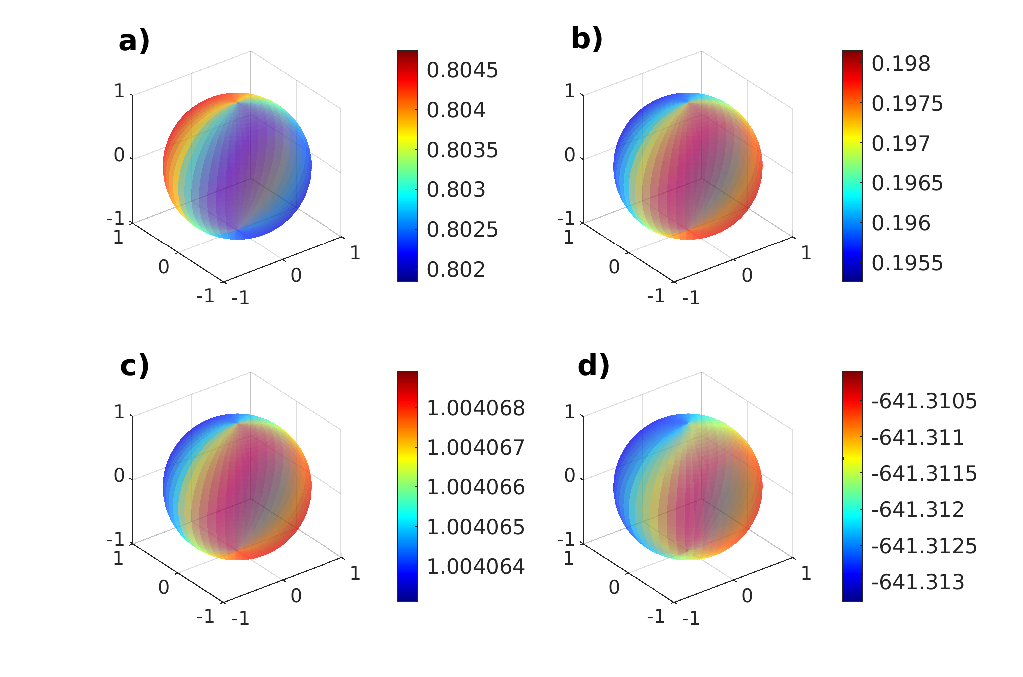}
    \caption{Surface fields. a) Dimensionless substrate concentration $\hat{C}_M = C_M/C_0$, b) Dimensionless product concentration $\hat{C}_N=C_N/C_0$, c) Dimensionless temperature $\hat{T}=T/T_0$, d) Dimensionless electrostatic potential $\hat{\psi}=\psi q_0/k_BT$. These results are obtained for $\alpha^2 = 10^{-2}$, $\beta^2=10^{-3}$, $\lambda^2 = 10^{-7}$, $\omega^2=10^{-5}$. Particle dimensions are also dimensionless, $\hat{r}=r/R$.}
    \label{fig:2D}
\end{figure}

In Fig.~\ref{fig:2D}(a), we observe that the substrate concentration is higher on the non-catalytic side (back of the sphere) compared to the catalytic side (front of the sphere), as expected. Conversely, in Fig.~\ref{fig:2D}(b) and (c), the highest product concentration and temperature are reached on the front side, where the reaction occurs, generating both product and heat. These figures highlight a stronger dependence on variations in $\phi$ than in $\theta$, with a clear symmetry around $\theta = \pi/2$. In Fig.~\ref{fig:2D}(d), the electric potential exhibits a behavior similar to that of the product concentration and temperature; however, near the boundary between the catalytic and non-catalytic regions, the variation in $\phi$ is smoother compared to the previous cases.

\begin{figure}[h!]
    \centering
    \includegraphics[width=0.95\linewidth]{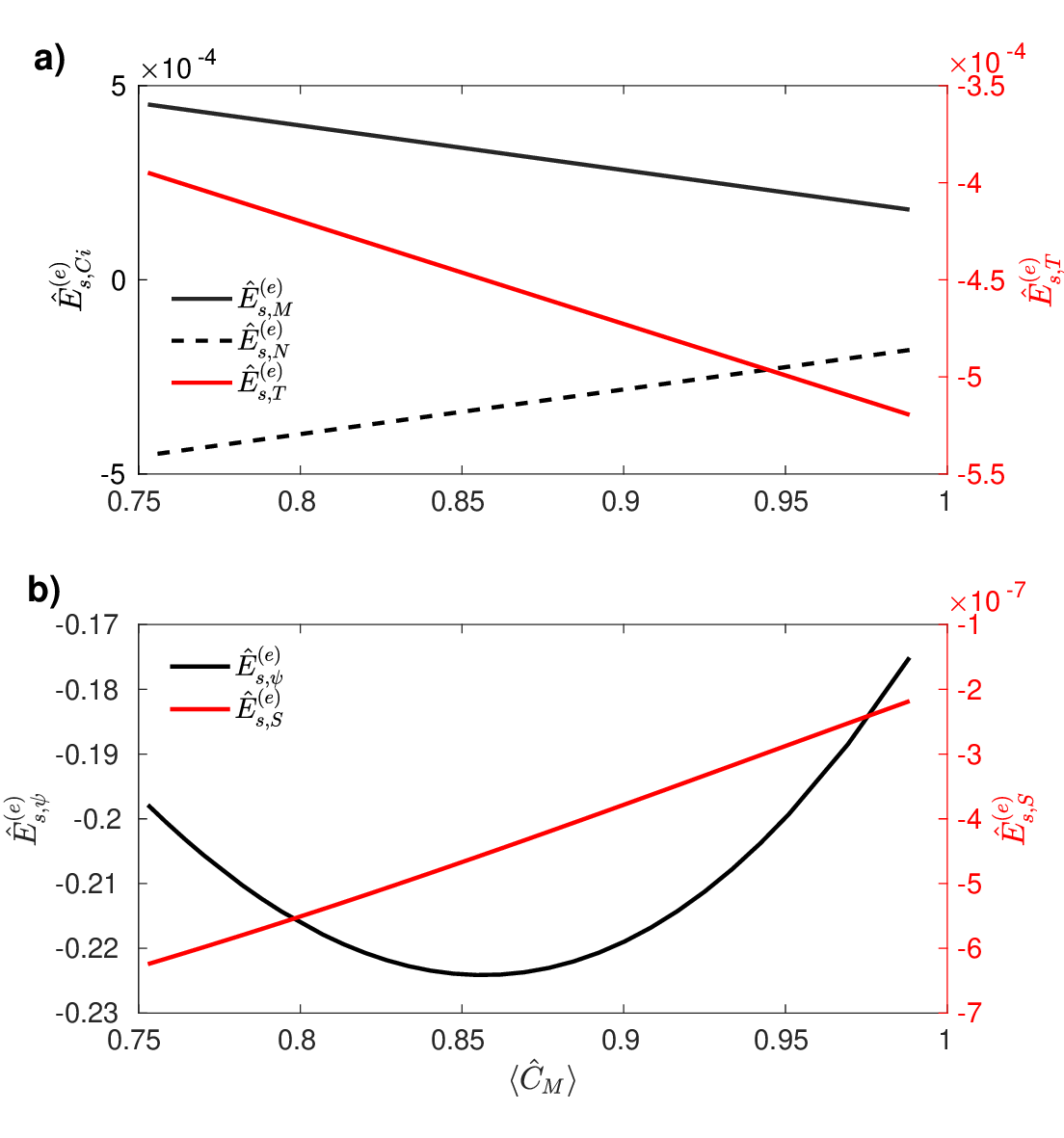}
    \caption{Dimensionless surface excess energy as a function of the mean substrate concentration.  
(a) Left y-axis: Contribution from variations in substrate and product concentrations:  
$\hat{E}_{s,C_i}^{(e)} = \frac{1}{4\pi R^2} \int_S \nabla_S \hat{C}_i \, dS$.  
Right y-axis: Contribution from temperature variations:  
$\hat{E}_{s,T}^{(e)} = \frac{1}{4\pi R^2} \int_S \nabla_S \hat{T} \, dS$. (b) Left y-axis: Contribution from the electrostatic potential:  
$\hat{E}_{s,\psi}^{(e)} = \frac{1}{4\pi R^2} \sum_i z_i \int_S \nabla_S \hat{\psi} \hat{C}_i \, dS$. Right y-axis:  Entropic contribution from variations in substrate and product concentrations $\hat{E}_{s,S}^{(e)} = \frac{1}{4\pi R^2} \sum_i \int_S \ln{\hat{C}_i}\nabla_S \hat{C}_i \, dS$   }
    \label{fig:Es_Dimless}
\end{figure}

In Fig.~\ref{fig:Es_Dimless}, we demonstrate the dependence of surface excess energy on the mean substrate concentration, as previously obtained in Table~\ref{tab:table2}. When considering only variations in product and substrate concentrations, as well as temperature, the dependence is linear with $\langle C_M \rangle$ (as shown in Table~\ref{tab:table2} for $E_{s,r}^{(e)}$, $E_{s,\gamma_{C_i}}^{(e)}$, and $E_{s,\gamma_T}^{(e)}$), see Fig. ~\ref{fig:Es_Dimless}(a). In contrast, when accounting for the effect of the electrostatic field, the dependence becomes quadratic with $\langle C_M \rangle$, as presented in Table~\ref{tab:table2} for $E_{s,\psi}^{(e)}$. In the case of the entropic effect, since the dimensionless concentration is far from zero, it is expected a linear behaviour as observed in Fig. ~\ref{fig:Es_Dimless}(b).
\\
In the context of the dissipative approach, Table~\ref{tab:table3} summarizes the contributions to the dissipative surface energy $E_s^{(d)}$. Specifically, $E_{s,S}^{(d)}$ corresponds to the entropic contribution from mixing, $E_{s,D}^{(d)}$ accounts for dissipation due to mass diffusion, $E_{s,Q}^{(d)}$ represents the contribution from heat conduction, $E_{s,r}^{(d)}$ is the dominant term associated with energy dissipation from the chemical reaction, and $E_{s,v}^{(d)}$ captures the dissipation due to surface velocity gradients. The table reveals both linear and quadratic dependencies on the average concentration. Among the terms, the entropy of mixing and the energy dissipated by the chemical reaction are expected to dominate, as both scale proportionally with $\alpha^2$.
\\
\begin{table}
\caption{\label{tab:table3}Dissipative surface energy contributions}
\begin{ruledtabular}
\begin{tabular}{ccc}
 Effect  & Result \\ \hline
 $E_{s,S}^{(d)}$&$-A\epsilon\tau k_r R_gT_0 \ln{\left(\langle \hat{C}_M\rangle(1-w_0)\right)}(\alpha^2+\beta^2)\langle C_M\rangle$\\
$E_{s,D}^{(d)}$&$A\varepsilon\frac{\tau D_MR_gT_0}{R^2}\beta^2\left(\frac{\pi^2}{24C_0}\beta^2\langle{C}_M\rangle + \frac{2}{3}\alpha^4\pi^2\right)\langle{C}_M\rangle$\\
$E_{s,Q}^{(d)}$&$A\varepsilon\frac{\tau D_M}{R^2}\alpha^2\beta^2\pi^2\frac{\lambda^2}{24\omega^2}\Delta H_r\langle{C}_M\rangle$ \\
$E_{s,r}^{(d)}$&$A\varepsilon\frac{\tau D_MR_gT_0}{R^2C_0}\frac{\sqrt{(\alpha/\beta)^2+1}^3}{\left( 1 + \sqrt{(\alpha/\beta)^2+1}^3\right)}\alpha^2\langle{C}_M  \rangle^2 $ \\
$E_{s,v}^{(d)}$&$\!\begin{aligned}[t]
    \frac{\varepsilon\tau}{\eta_s}&\left[ \beta^4\gamma_{C_M}^2 + \alpha^4\gamma_{C_N}^2 + \lambda^4\frac{T_0^2}{C_0^2}\gamma_T^2 \right. \\& \left.+\left(\frac{Fk_BTR}{q_0}\right)^2\left( z_M^2 + 2\alpha^2z_Mz_N + \alpha^4z_N^2\right) \right]\langle{C}_M\rangle^2\end{aligned}$ \\
\end{tabular}
\end{ruledtabular}
\end{table}

\section{Results and Discussion}

To validate our theoretical framework, we analyze two experimentally studied systems where active diffusivity has been observed to depend on the reaction rate. The first case considers nanometric catalytic Janus particles that utilize a charged salt as a substrate \cite{Qin2017_self-thermo}, while the second examines phospholipid vesicles with embedded enzymes hydrolyzing ATP \cite{Ghosh2019}. To obtain the results presented in Fig. \ref{fig:nano} and Fig. \ref{fig:Liposomes}, we used the experimental parameter values ($R$, $T_0$, $k_r$, $z_M$, $z_N$, $\Delta H_r$, $\Delta \mu_r$) provided in Refs. \cite{Qin2017_self-thermo,Ghosh2019}. The dependencies of surface tension on concentration and temperature ($\gamma_{C_M}$, $\gamma_{C_N}$, $\gamma_{T}$), as well as diffusivities, were extracted from the literature \cite{LI2001,mehringer2021hofmeister,gordon1966role,hubley1996effects}. Thermal conductivity, electric permittivity, and interface thickness, were estimated using data of Refs.\cite{LI2001,mehringer2021hofmeister,gordon1966role,hubley1996effects}. Finally, to vary the average concentration and, consequently, the average reaction rate, as reported in experiments, we varied the substrate bulk concentration.  
\\
We demonstrate that the increase in active diffusivity is a direct consequence of the rise in surface excess energy with the reaction rate. Crucially, we identify the average surface substrate concentration as the key parameter governing this effect. Furthermore, our analysis shows that surface excess energy is influenced not only by the reaction rate but also by the reaction heat, affinity, particle size, surface tension variations, and the charge of chemical compounds at the interface. This comprehensive understanding allows us to bridge experimental observations with a theoretical foundation, offering new insights into the underlying mechanisms driving enhanced mobility in active systems.
\\
The parameter values used to obtain our model results are: $R = 9$nm, $T_0=298$K, $k_r=21.4$s$^{-1}$, $z_M=-3$,$z_N=-4$,$\Delta H_r=-91$kJ/mol, $\Delta \mu_r = -30$kJ/mol, $U=10$s$^{-1}$, $U_q = 0.02$m/s, $\kappa=0.56$W/mK, $\gamma_{C_M} = 1.5$mJ/m$^2$M, $\gamma_{C_N} = 0.2$mJ/m$^2$M, $\gamma_{T} = -0.2$mJ/m$^2$K, $\varepsilon = 5$nm, and therefore $\alpha^2 = 1\times10^{-5}$, $\beta^2 = 4.98\times10^{-6}$, $\omega^2 = 1\times10^{-18}$, for the case of the nanometric catalytic Janus particle. Regarding the vesicles with embedded enzymes hydrolyzing ATP, we assume a non-homogeneous distribution of the enzymes, most of them on one side of the vesicle. The parameters considered for the vesicle diffusivity are $R = 2$$\mu$m, $T_0=294$K, $k_r=25$s$^{-1}$, $z_M=0$,$z_N=-1$,$\Delta H_r=-30$kJ/mol, $\Delta \mu_r = -60$kJ/mol, $U=10$s$^{-1}$, $U_q = 1$m/s, $\kappa=0.56$W/mK, $\gamma_{C_M} = -0.01$mJ/m$^2$M, $\gamma_{C_N} = 0.02$mJ/m$^2$M, $\gamma_{T} = -0.01$mJ/m$^2$M, $\varepsilon = 15$nm, and therefore $\alpha^2 = 0.31$, $\beta^2 =0.12$, $\omega^2 = 8\times10^{-12}$.

\subsection{Nanometric catalytic Janus particles}
\begin{figure}
    \centering
    \includegraphics[width=0.9\linewidth]{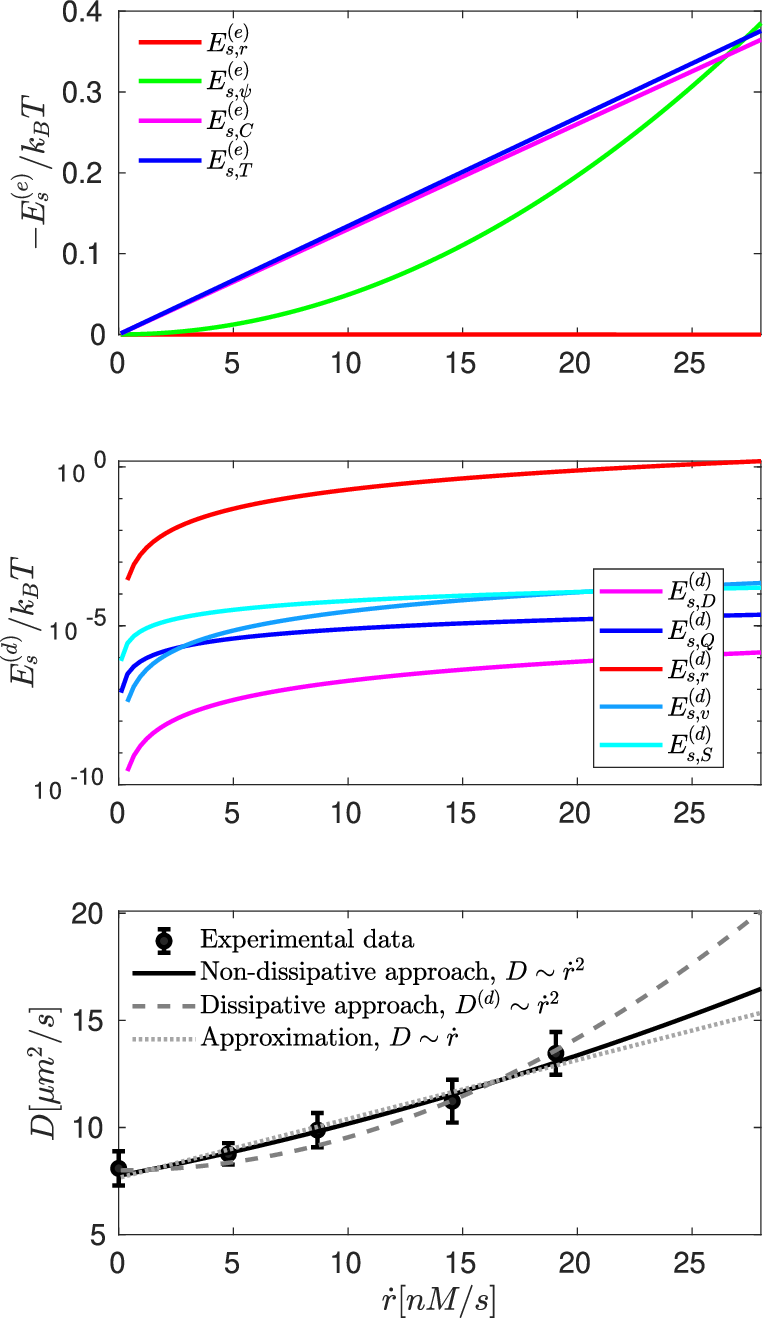}
    \caption{Non-dissipative and dissipative surface excess energy and active diffusivity of nanometric Janus particles as a function of the average reaction rate $\dot{r} = k_r\langle C_M \rangle$[nM/s]. (a) Negative surface excess energy $-E_s^{(e)}/k_BT$  computed from Eq.(\ref{Excess}). (b) Dissipative surface excess energy $E_{s}^{(d)}/k_BT$  computed from Eq.(\ref{Diss_Excess}). (c) Active diffusivity $D$ computed from Eq.(\ref{Diff}) (continuous black line), active diffusivity computed from dissipative approach $D^{(d)}$ from Eq. (\ref{D_d}) (dashed dark grey line) whereas the black dots with the error bars represent the experimental data \cite{Qin2017_self-thermo}, all following a quadratic dependence with $\dot{r}$. The dotted light grey line corresponds to the approximation proposed in Ref. \cite{Qin2017_self-thermo}, which considers only the self-thermophoretic effect, and follows a linear dependence on $\dot{r}$. The model results are shown for $1\times10^{-19}\le\lambda^2\le1.13\times10^{-17}$, $3\times10^{-4}\le\xi^2\le3.29\times10^{-2}$ and $\tau = (k_r\beta^2)^{-1}$.}
    \label{fig:nano}
\end{figure}

In Fig. \ref{fig:nano}, we present the surface excess energy and active diffusivity of nanometric Janus particles. Fig. \ref{fig:nano}(a) illustrates the behavior and relative contributions of the different terms in the surface excess energy. The blue line represents the self-thermophoretic contribution $E_{s,T}^{(e)}$, the magenta line corresponds to the self-diffusiophoretic term $E_{s,\gamma_C}^{(e)}$, the green line depicts the self-electrophoretic term $E_{s,\psi}^{(e)}$, and the red line represents the reaction-phoretic contribution $E_{s,r}^{(e)}$. We observe that both self-thermophoretic and self-diffusiophoretic terms dominate, exhibiting similar trends and magnitudes. Consequently, the enhancement of active diffusivity (Fig. \ref{fig:nano}(c)) is not solely due to self-thermophoretic effects, as previously suggested in Ref. \cite{Qin2017_self-thermo}, but also significantly influenced by self-diffusiophoretic contributions. 
\\
In Fig.~\ref{fig:nano}(b), we present the contributions to the dissipative surface excess energy on a logarithmic scale. It is evident that the energy dissipation associated with the chemical reaction, $E_{s,r}^{(d)}$ (red line)—a scalar process—dominates at the nanoscale. In contrast, the contributions from vectorial and tensorial processes, such as mass diffusion $E_{s,D}^{(d)}$ (magenta line), heat conduction $E_{s,Q}^{(d)}$ (blue line), and fluid flow $E_{s,v}^{(d)}$, are negligible. Additionally, the entropic contribution from mixing, $E_{s,S}^{(d)}$, is also minor compared to the entropy produced by irreversible processes. These results indicate that, from a dissipative perspective, the enhancement of diffusivity observed in Fig.~\ref{fig:nano}(c) is primarily driven by the energy dissipation associated with the chemical reaction occurring at the nanoparticle surface.
\\
In Fig. \ref{fig:nano}(c), the continuous black lines and dashed dark grey line shows our theoretical predictions from the non-dissipative and dissipative approaches respectively following a quadratic dependence on the reaction rate whereas the light grey dotted line shows a linear approximation \cite{Qin2017_self-thermo}, when considering $\gamma_{T} = -0.5$mJ/m$^2$K, three orders of magnitude larger than literature estimates. The black dots with error bars correspond to experimental data from Ref. \cite{Qin2017_self-thermo}, which follows a quadratic behaviour (with $R_r = 0.993$). We observe that self-electrophoresis plays a crucial role, scaling quadratically with the average reaction rate and improving agreement between model results and experimental data. Notably, self-thermophoresis alone cannot fully explain the experimental results, as it predicts a linear dependence of $D$ on $\dot{r}$ and requires a $\gamma_T$ value far from physical meaning. Regarding the dissipative approach, it is worth noting that it captures the quadratic dependence on the reaction rate and provides a sufficiently accurate fit to the experimental data when considering the characteristic time $\tau = (k_r \beta^2)^{-1}$. This implies that, at the nanoscale, the relevant timescales must account for the interplay between reaction kinetics and diffusion along the interface. The key insight is that the energy dissipated during the chemical reaction sustains the gradients required to increase the surface energy, thereby enhancing the particle's diffusivity.
\\
{The observed discrepancy between the dissipative and non-dissipative approaches, particularly at high reaction rates, indicates that computing the dissipative surface excess energy, for nanometric particles, solely from entropy changes—both reversible and irreversible—may not fully capture the system’s behavior. This suggests that entropy production and mixture entropy alone are insufficient to describe the dynamics at large activities for such small systems. Under these conditions, additional non-dissipative contributions, such as those captured by the concept of frenesy, become important\cite{MAES20201}. A comprehensive description should therefore combine both dissipative contributions (entropy changes) and non-dissipative effects (frenesy), providing a unified framework capable of explaining the system’s behavior across the full range of activities.}

\subsection{Phospholipid vesicles with embedded enzymes}

\begin{figure}[b]
    \centering
    \includegraphics[width=0.9\linewidth]{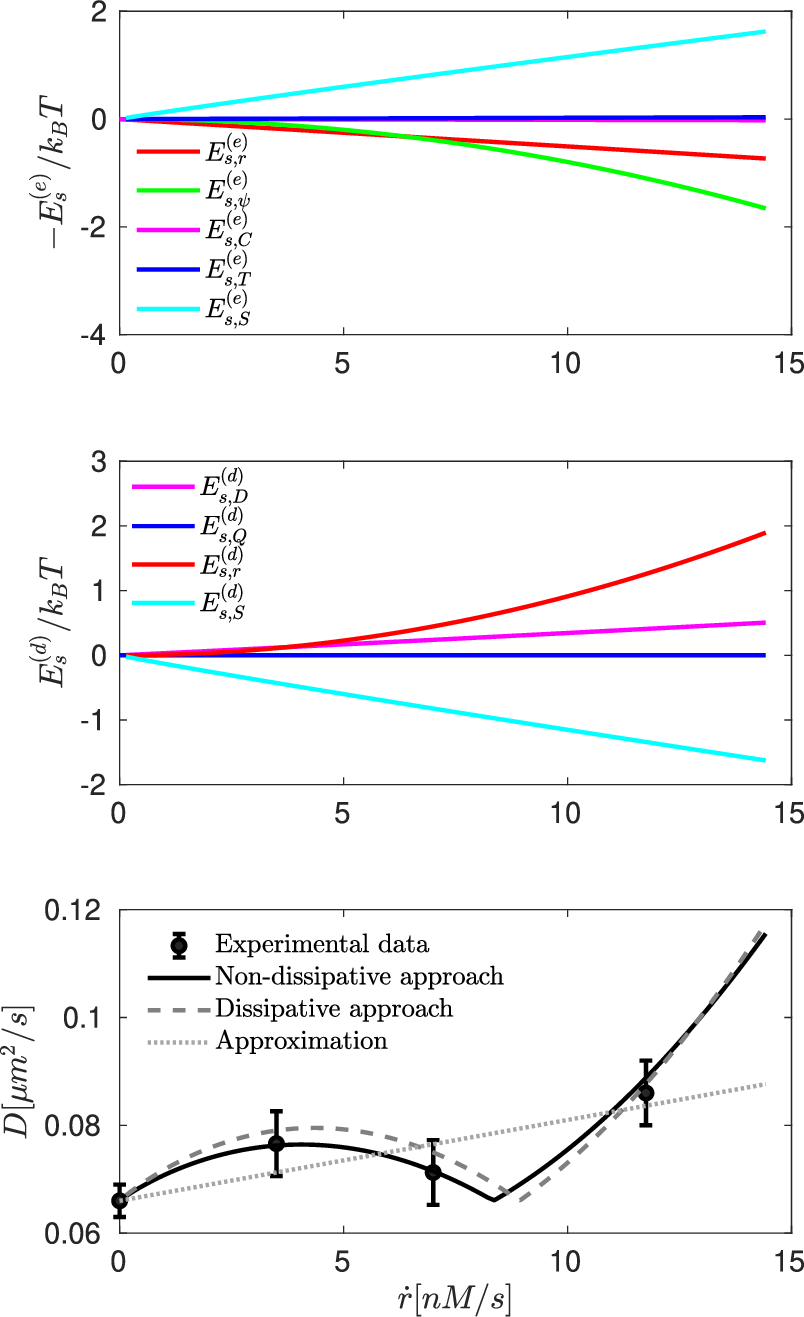}
    \caption{Non-dissipative and dissipative surface excess energy and active diffusivity of catalytic liposomes as a function of the average reaction rate $\dot{r} = k_r\langle C_M \rangle$[nM/s]. (a) Negative surface excess energy $-E_s^{(e)}$[J] computed from Eq.(\ref{Excess}). (b) Dissipative surface excess energy $E_{s}^{(d)}/k_BT$  computed from Eq.(\ref{Diss_Excess}).(c) Active diffusivity $D$ computed from Eq.(\ref{Diff}) (continuous black line), active diffusivity computed from dissipative approach $D^{(d)}$ from Eq. (\ref{D_d}) (dashed dark grey line) whereas the black dots with error bars correspond to experimental data from Ref.\cite{Ghosh2019}. The dotted light grey line corresponds to a linear approximation without considering entropic effects. For $1\times10^{-15}\le\lambda^2\le2\times10^{-13}$,$3\times10^{-2}\le\xi^2\le2.9$ and $\tau=k_r^{-1}$.}
    \label{fig:Liposomes}
\end{figure}

In Fig.~\ref{fig:Liposomes}, we present the surface excess energy and active diffusivity of phospholipid vesicles with embedded enzymes. Fig.~\ref{fig:Liposomes}(a) illustrates the behavior and relative contributions of the different terms in the surface excess energy. The blue line represents the self-thermophoretic contribution $E_{s,T}^{(e)}$, the magenta line corresponds to the self-diffusiophoretic term $E_{s,\gamma_C}^{(e)}$, the green line depicts the self-electrophoretic term $E_{s,\psi}^{(e)}$, the red line represents the reaction-phoretic contribution $E_{s,r}^{(e)}$, and the cian line corresponds to the entropic contribution $E_{s,S}^{(e)}$. Unlike the previous case of the nanometric catalytic Janus, entropic contributions become significant due to the considerable increase in particle size, increasing $\alpha^2$ and $\beta^2$.  On the other hand, both self-thermophoretic and self-diffusiophoretic contributions are negligible, as the surface tension exhibits weak dependence on substrate and product concentrations as well as temperature. Consequently, entropic, enthalpic, free energy, and self-electrophoretic effects dominate and compete among them, leading to a non-monotonic behavior of active diffusivity $D$ as a function of the mean reaction rate (Fig.~\ref{fig:Liposomes}(c)). {This non-monotonicity arises because the self-electrophoretic and reaction-phoretic contributions can become negative depending on parameters such as the charge ($z_M$ and $z_N$) and concentration of reactants and products ($C_M$ and $C_N$), the signs of the derivatives of surface tension with respect to temperature ($\gamma_T$) and concentration ($\gamma_{C_M}$ and $\gamma_{C_N}$), and the sign of the reaction’s free energy change ($\Delta \mu_r$). Variations in these parameters modulate the relative magnitude and sign of each contribution, producing the observed complex dependence of $D$ on the mean reaction rate.}
\\
In Fig.~\ref{fig:Liposomes}(b), we show the contributions to the dissipative surface excess energy. In this case, the entropic contribution from mixing, $E_{s,S}^{(d)}$ (cyan line), and the dissipative contributions from the chemical reaction, $E_{s,r}^{(d)}$ (red line), and mass diffusion, $E_{s,D}^{(d)}$ (magenta line), are found to be of comparable magnitude. This results in a competition among these effects, giving rise to the non-monotonic behavior of $D^{(d)}$ observed in Fig.~\ref{fig:Liposomes}(c). Both the dissipative and non-dissipative approaches lead to the same active diffusion behaviour.
\\
In Fig.~\ref{fig:Liposomes}(c) the continuous black line and dashed dark grey line show our theoretical predictions, whereas the dotted light grey line shows a linear approximation when considering $\gamma_{C_M} = -1$mJ/m$^2$K or $\gamma_{C_M} = 1$mJ/m$^2$K and neglecting entropic effects. The black dots with error bars correspond to experimental data from Ref.\cite{Ghosh2019}. Notably, around $\dot{r} = 8.1$~nM/s, the active diffusivity reaches its minimum value, corresponding to the point where the surface excess energy cancels. {The small shift observed in Fig. 5(c) between the non-dissipative and dissipative approaches arises from the value of the dimensionless time $\tau = k_r^{-1}$, which may differ slightly from the experimental value.} 
\\
We observe that the enhancement of active diffusivity (Fig.~\ref{fig:Liposomes}(c)) is not driven by self-diffusiophoretic effects, as previously suggested in Ref.~\cite{Ghosh2019}, but rather by self-electrophoretic and reaction-phoretic contributions, or, from the energy dissipated by the chemical reaction. Moreover, self-electrophoresis remains significant, as it scales quadratically with the average reaction rate, leading to improved agreement with experimental results, (see Fig.~\ref{fig:Liposomes}(c)). {The entropy contribution in Fig. ~\ref{fig:nano}(a) and the fluid flow contribution in Fig. ~\ref{fig:Liposomes}(b) are not shown because they are negligible compared to the other contributions; we exclude them to keep the figures uncluttered.}

\section{Conclusions}
In summary, we have demonstrated that the active diffusivity of catalytic nanoparticles and enzyme-functionalized vesicles depends on the excess surface energy generated by the chemical reaction. This excess surface energy was estimated using two complementary approaches: a non-dissipative approach, based on the difference between internal and surface energy, and a dissipative approach, focused on the energy dissipated at the interface due to irreversible processes.
\\
For Janus particles, we observe that self-thermophoresis and self-diffusophoresis contribute comparably to active motion, while self-electrophoresis introduces a quadratic dependence on reaction rate, which better fits experimental observations. From the dissipative perspective, this quadratic behavior is also captured and explained by the dissipation of energy derived from the chemical reaction at the surface.
\\
In the case of enzyme-functionalized vesicles, the situation is more complex. Competing effects, including variations in surface tension due to mixing entropy, the electrostatic field, and reaction heat, lead to a complex response in which self-diffusiophoresis and self-thermophoresis are negligible. Dissipative analysis also reveals a competition between the contribution of mixing entropy, which is quasi-linear with reaction rate, and entropy production due to irreversibility, which is predominantly quadratic with reaction rate, reinforcing the non-monotonic nature of particle diffusivity as a function of reaction rate observed in experiments.
\\
Together, these results show that both dissipative and non-dissipative approaches converge on consistent predictions of active diffusivity observed experimentally, supporting the idea that the concepts of frenesy\cite{MAES20201} and self-organization driven by energy dissipation \cite{Arango-Restrepo2021} provide complementary and robust tools for understanding anomalous transport in non-equilibrium systems.
\\
Our analysis highlights that excess surface energy is fundamental to explaining active diffusion. Mechanisms such as self-diffusion, self-thermophoresis, self-electrophoresis, reaction-phoresis, and entropic effects can compete or act synergistically to modulate mobility, while energy dissipation due to entropy production provides a unifying framework to holistically capture their combined effect. We emphasize that modeling active diffusivity requires considering the coupling between multiple processes rather than treating them as isolated contributions.
\\
Finally, we conclude that the improvement in mobility stems from variations in surface tension maintained by entropy production from interfacial chemical reactions. The excess surface energy generated by exothermic reactions emerges as a key factor in active transport, offering new insights into chemically driven motion and guiding the design of synthetic microswimers and active colloidal systems.

\begin{acknowledgments}
The authors are grateful for the financial support of MICIU (Spanish Government) under grant No. PID2021-126570NB-I00.
\end{acknowledgments}
\section*{Data Availability Statement}
Data available in the manuscript or supplementary material. The data that support the findings of this study are available within the article.

\appendix

\section{Orientation Correlation Function in the Overdamped Limit}

We consider a particle with orientation vector \( \mathbf{n}(t) \), evolving under stochastic rotational dynamics in the absence of active torques. The overdamped Langevin equation for the orientation is given by Eq.(\ref{omega}), where the angular velocity is modeled as Gaussian white noise in the overdamped limit $\langle \omega_i(t) \omega_j(t') \rangle = 2 D_r \delta_{ij} \delta(t - t')$, with \( D_r \) the rotational diffusion coefficient.
\\
Over time, the particle orientation vector undergoes random rotations due to stochastic torques. The dynamics correspond to a rotational diffusion process, for which the orientation correlation decays exponentially:
\begin{equation}
    \langle \mathbf{n}(t) \cdot \mathbf{n}(t') \rangle = e^{-2 D_r |t - t'|}.
\end{equation}
This is valid in the limit where inertial effects are negligible and the particle's orientation undergoes a Markovian random walk on the particle\cite{Debye1929}.
\\
Since the system is isotropic (no preferred direction), we can compute the tensorial two-time correlation function \cite{Chandler_I1974,Chandler_II1974}:
\begin{equation}
    \left\langle n_i(t) n_j(t') \right\rangle = \left[ \frac{1}{3} + \left( \langle \mathbf{n}(t) \cdot \mathbf{n}(t') \rangle - \frac{1}{3} \right) \right] \delta_{ij}.
\end{equation}
{Regarding the phoretic torque, in the absence of a preferred direction the self-induced local fields either align with a unit director vector, $\mathbf{m}\to\mathbf{n}$ (when the rotational relaxation time is sufficiently short for a gradient to develop), or remain completely random, $\mathbf{m}$, such that $\psi \to 0$. In both cases, the phoretic torque vanishes, either $\mathbf{T}{ph}\to 0$ at all times, or $\langle \mathbf{T}{ph}\rangle \to 0$ over long times. Consequently, $|\mathbf{T}{ph}|\ll |\mathbf{T}{r}|$ at long times, and the phoretic torque becomes negligible compared to the random torque.}

\section{Mean translational kinetic energy}
The solution of the Langevin equation (Eq.(\ref{lang_t})) in 1-D is
\begin{equation}
\label{v_sln}
    {v}(t) = e^{-\xi_t t/m} {v}(0)+ \int_{0}^{t} dt' e^{-\xi_t (t-t')/m} ({F_{ph}}(t') + {F_{t}}(t'))
\end{equation}
Multiplying this expression by $ {v}(t)$ and taking the ensemble average, we obtain:
\begin{equation}
\label{A2}
\begin{split}
     \langle {v}(t)^2 \rangle  = & e^{-2\xi_t t/m} {v}(0)^2+ \\&\frac{1}{m^2}\int_{0}^{t} dt' e^{-\xi_t (t-t')/m} \int_{0}^{t} dt'' e^{-\xi_t (t-t'')/m} {H}
\end{split}
\end{equation}
in which ${H} = \langle({F_{ph}}(t') + {F_{t}}(t'))({F_{ph}}(t'') + {F_{t}}(t''))\rangle$ and have used the fact that $\langle{F_{ph}}(t') + {F_{t}}(t')\rangle = \langle{F_{ph}}(t'') \rangle + \langle {F_{t}}(t'')\rangle = 0$, since the noise sources are not correlated {because stochasticity of the phoretic force originates from the orientation vector $\mathbf{n}$, governed by Eq. (\ref{omega}), with a rotational noise source $\mathbf{T}_r$  independent of the translational noise source $\mathbf{F}_t$}. Using this fact again, we have 
\begin{equation}
\begin{split}
    H &=  \langle{F_{ph}}(t'){F_{ph}}(t'') \rangle + \langle {F_{t}}(t'){F_{t}}(t'')\rangle \\& = 2\xi_tB + 2\xi_t k_BT
    \end{split}
\end{equation}
Substituting this expression in Eq.(\ref{A2}), and solving the integral, we obtain:
\begin{equation}
   \langle {v}(t)^2 \rangle  = e^{-2\xi_t t/m} {v}(0)^2+ 2\frac{B + k_BT}{m}\left( 1 - e^{-2\xi_t t/m}\right)
\end{equation}
The mean squared velocity at equilibrium ($t\to\infty$) is then the sum of the phoretic and thermal contributions
\begin{equation}
   \langle {v}(t)^2 \rangle_{t\to\infty}  =  2\frac{B}{m} + 2\frac{k_BT}{m}
\end{equation}
The mean translational kinetic energy of the particle is then
\begin{equation}
\label{kin}
  \frac{m}{2}\langle{v}(t)^2 \rangle_{t\to\infty}  =  B + k_BT 
\end{equation}
and by considering that the surface excess energy $\left|E_s^{(e)}\right|$ is the source of energy of the phoretic force, we find that $B = \left|E_s^{(e)}\right|$. 

\section{Dimensionless balance equations}

The dimensionless substrate and product concentration at the particle interface are given by:
\begin{equation}
\label{mass_eq_s}
    0 =\frac{\partial^2 \hat{C}_{M}}{\partial \phi^2} - \alpha^2\hat{C}_{M} \Theta(\phi_0-\phi) - \beta^2(\hat{C}_{M}-\hat{C}_{M}^{(b)})
\end{equation}
where $\alpha^2 = k_{r}R^2/D_s$ and $\beta^2 = UR^2/D_s$ are the dimensionless numbers quantifying the reaction-diffusion and adsorption-diffusion effects on the interface. For the substrate we have
\begin{equation}
\label{mass_eq_p}
    0 =\frac{\partial^2 \hat{C}_{N}}{\partial \phi^2} + \alpha^2\hat{C}_{M} \Theta(\phi_0-\phi) - \beta^2\hat{C}_{N}
\end{equation}
We have assumed that the product concentration at the interface greatly exceeds its concentration in the bulk.
\\
On the other hand, the dimensionless temperature at the interface fulfills the equation
\begin{equation}
\label{ene_eq}
   0 = \frac{\partial^2\hat{T}}{\partial \phi^2} +   \Theta(\phi_0-\phi)\lambda^2\hat{C}_{M} - \omega^2(\hat{T}-\hat{T}^{(b)})
\end{equation}
in which for an exothermic reaction, $\lambda^2 = \frac{R^2|\Delta H_{r}| k_{r}C_{0}}{kT_{0}}$, and $\omega^2=\frac{U_{q}R}{k}$. These dimensionless numbers represent the effects of heat generation-conduction and cooling-conduction, respectively. 
\\
When substrate and product are charged species, the dimensionless Poisson-Boltzmann equation is
\begin{equation}
    \frac{\partial^2\hat{\psi}}{\partial\phi^2} = - \xi^2\sum_i z_i\hat{C}_i
\end{equation}
with $\xi^2 = \frac{R^2C_0N_Aq_0^2}{k_BT\varepsilon}$, and $\hat{\psi}=\psi q_0/k_BT$.

 \bibliographystyle{IEEEtran}
 \bibliography{Manuscript}
 
\end{document}